\documentclass[10pt,letterpaper]{article}
\usepackage[utf8]{inputenc}
\usepackage{amsmath}
\usepackage{amsfonts}
\usepackage{amssymb}
\usepackage{graphicx}
\usepackage{color}
\usepackage[normalem]{ulem}
\usepackage[left=2cm,right=2cm,top=2cm,bottom=2cm]{geometry}

\newtheorem{theorem}{Theorem}[section]
\newtheorem{lemma}{Lemma}[section]

\newtheorem{definition}{Definition}[section]

\newcommand{\qed}{\hfill$\Box$\par\medskip}

\def\bhag#1{\noindent
\setcounter{equation}{0}
\section{#1}
}

\def\RR{{\mathbb R}}
\def\CC{{\mathbb C}}
\def\ZZ{{\mathbb Z}}

\def\PPI{{{\rm I}\kern-1pt\Pi}}

\def\a{\alpha}
\def\b #1;{{\bf #1}}
\def\x{{\bf x}}
\def\k{{\bf k}}
\def\y{{\bf y}}
\def\u{\mathbf{u}}
\def\w{{\bf w}}
\def\z{{\bf z}}

\def\j{\mathbf{j}}

\def\be{\begin{equation}}
\def\ee{\end{equation}}
\def\bea{\begin{eqnarray}}
\def\eea{\end{eqnarray}}
\def\eref#1{(\ref{#1})}

\def\donchitre#1#2{\vskip 6.5cm\noindent
\parbox[t]{1in}{\special{eps:#1.eps x=6.5cm y=5.5cm}}
\hbox to 7cm{}\parbox[t]{0.0cm}{\special{eps:#2.eps x=6.5cm y=5.5cm}}}

\def\tn{|\!|\!|}

\def\bs#1{{\boldsymbol{#1}}}

\def\ns{\tilde{\tau}}

\begin{document}
\title{A Fourier-invariant method for locating point-masses and computing their attributes}
\author{Charles K. Chui\\
Department of Mathematics, Hong Kong Baptist University, Hong Kong 
\thanks{This author is also associated with the Statistics Department of Stanford University, CA 94305, and his research is partially supported by U.S. ARO Grant W911NF-15-1-0385.}\\
\textsf{email:} ckchui@stanford.edu.  \\
H.~N.~Mhaskar\\
Institute of Mathematical Sciences, Claremont Graduate University,\\ Claremont, CA 91711.
\thanks {The research of this author is supported in part by ARO Grant W911NF-15-1-0385.}\\
\textsf{email:} hrushikesh.mhaskar@cgu.edu. }

\date{}

\maketitle

\begin{abstract}
Motivated by the interest of observing the growth of cancer cells among normal living cells and exploring how galaxies and stars are truly formed, the objective of this paper is to introduce a rigorous and effective method for counting point-masses, determining their spatial locations, and computing their attributes. Based on computation of Hermite moments that are Fourier-invariant, our approach facilitates the processing of both spatial and Fourier data in any dimension.  
\end{abstract}
                           
\bhag{Introduction}\label{intsect}

Throughout the history of mankind, people have always been fascinated by twinkling stars in the sky, and many have tried but failed to count them. The number of stars visible to the naked eye is estimated to be over $6,000$, but one can see far less than half of them at the same time, even on a clear night, from either the northern or southern hemisphere. 
The question of how many stars in the universe has fascinated scientists, philosophers, poets, musicians, and dreamers throughout the ages. Ever since the invention of the telescope, commonly credited to the Dutch lens-maker, Hans Lippershey, in 1608, but significantly modified and improved by the “father of observational astronomy”, Galileo Galilei, less than a year later, many types of powerful telescopes have been invented, particularly in the twentieth century, with the capability of observing celestial objects beyond the (human) visible light (of wavelengths from about 390 nm to 700 nm), and ranging from the long radio wavelengths to the short wavelengths of gamma rays. 
However, though not being the very first, the launch of the Hubble space telescope (HST), on April 24, 1990, was surely a major breakthrough in (observational) astronomy. Indeed, HST allows scientists to observe stars (and even some of their planets and moons) in the electromagnetic spectrum that ranges between near-ultra violet and near-infrared, from outer space to avoid atmospheric turbulence effects that cause the  “twinkling” visual and other artifacts, when observing stars from ground-based telescopes. While the HST orbits the Earth at an altitude of about 353 miles, the space telescope, Spitzer, launched in 2003, was designed to trail the Earth orbit and to scan the skies in infrared. The “image data” gathered by Hubble and Spitzer together, perhaps along with other space telescopes as well, allow us to have an approximate count of $70$ billion trillion (or $7\times 10^{22}$) stars and over $100$ billion galaxies in the universe, as well as more than $300$ billion stars in our own Milky Way (galaxy) alone.

The problem of observing, counting, isolating, and tracking (perhaps moving) point-masses is not limited to the investigation in astronomy. It is also important for research in biomedical sciences. For the heath-care sector, counting (red, white, and platelet) blood cells is routine in any regular physical examination. The importance of counting red blood cells (RBCs), known as erythrocyte count, is that the number of RBCs can affect the amount of oxygen the tissues receive, since RBCs contain hemoglobin which carries oxygen to the tissues. On the other hand, counting white blood cells (WBCs), along with their differentials (that is, breakdown of percentage of each WBC type) is for assessing the ability of the body to fight infection, since WBCs are produced and derived from multipoint cells in the bone marrow for protecting the body against both infectious disease and foreign invaders. The major function of platelets is to prevent bleeding, and its count is perhaps less important, since platelets are very small and make up only a tiny fraction of the blood volume anyway. The current practice in regular physical examination is to draw a test tube of blood from the patient for analysis in the laboratory. An alternative is to take a small blood sample, by drawing off a finger prick with a “Pasteur pipette”, for immediate processing by an automatic counter. However, though less invasive, more timely, and low-cost, this alternative is usually not quite reliable.  For non-invasive health care practice, it is clear that blood cell counting from digital imaging would be much more desirable. With the recent rapid advancement of biosensor and super-resolution optical microscopy technologies, it makes sense to invest more research effort in this direction. In this regard, various medical devices are already or soon to be available, and there are research papers, reports, and U.S. patents, such as \cite{roy2014simple, chung2015counting, fang2017chip,li2015red, lou2016automatic, wegel2016imaging,  huang2016ultra, GEpaidad}, in the literature or web-pages. However, to the best our knowledge, there is no mathematically rigorous study in the current literature on counting RBCs and WBCs (along with their differentials) from imagery data without observation under the microscope.

The objective of this paper is to introduce a rigorous and effective method for  counting,  locating,  and isolating  point-masses along with their attributes, such as: galaxies and stars from spectral data acquired by space telescopes; red and white blood cells or other living cells from super-resolution optical microscopy; and other scientific applications. Before giving a precise formulation of the mathematical problem, let us first consider the ideal case, where the point-masses are represented by points (with zero measure), there is no noise nor other associated contamination or perturbation, and if applicable, the optical lens is an ideal thin lens that provides a Fourier transform optical observation tool. In other words, to motivate the statement of the mathematical problem, let the point-masses along with their attributes be represented by the counting measure:
\be\label{originalmeasure}  
\tau_I =\sum_{\ell=1}^L a_\ell\delta_{\x_\ell},  
\ee
where $\delta_{\x_\ell}$ denotes the Dirac delta at $\x_{\ell}$, and both the number of terms, $L$, in the summation, as well as the locations $\x_{\ell}$ along with their coefficients $a_{\ell}$ are unknown. Hence, an equivalent formulation of $\tau_I$, via observation using the ideal thin lens, is given by:
\be\label{fourtrans}
  f_I(\u) = \sum_{\ell=1}^L a_\ell \exp(-i\u\cdot\x_\ell), 
\ee
which is an exponential sum. Although point-evaluation samples from $f_I$ would seem to a natural approach to finding $L$ and computing ${\x_\ell}$ and $a_\ell$, for $\ell=1, \cdots, L$, by applying an appropriate version of the Prony method, the difficulty is that $\x_\ell$ are points in the $q$-dimensional space $\RR^q$, with $q \ge 2$, and some of them may be very very close to one another. Furthermore, as to be discussed in the next section, we are interested in the realistic setting, where noise and/or perturbation are present. Therefore, instead of point-evaluation samples (that do not make sense for the counting measure $\tau_I$ anyway), we will consider data samples of integral moments in this paper. 

By considering moments $\widehat{\tau_I}(\k)$ with respect to the Hermite functions $\psi_\k(\x)$ (see \eref{hermitegenfunc} and \eref{hermitemoments} for the definition), it happens that moment data from both $\tau_I$ and $f_I$ are essentially the same, since Hermite functions are invariant under Fourier transformation, up to a multiplication constant (see \eref{fourhermitemoment}). In other words, such (moment) data samples 
$$
\widehat{\tau_I}(\k)= \sum_{\ell=1}^L a_\ell \psi_\k(\x_\ell), 
$$
acquired from either the spatial or frequency domains, can be used to determine the number $L$ as well as to compute the coordinates of the points $\x_\ell$ and the coefficients $a_\ell$, which are attributes of the corresponding point-masses, represented by the points  $\x_\ell \in\RR^q$. Since the family of functions $\{\psi_\k\}$ is a complete orthonormal set in $L^2(\RR^q)$, for any $q\ge 1$, the moment information in \eref{hermitemoments} is sufficient to determine the measure $\tau_I$ uniquely. In practice, we need to work with only finitely many of these moments. For this purpose, we will introduce the \textbf {point-mass isolation operator} or  \textbf {point-mass isolation operation}, collectively denoted by PIO and defined by
$$
\mathcal{T}_n(\{\widehat{\tau_I}(\k)\}; \x):=n^{-q}\sum_{\j\in\ZZ^q_+}H(\sqrt{|\j|_1}/n)\widehat{\tau_I}(\j)\psi_\j(\x),
$$ 
where $q \ge 1$, depending on the spatial dimension of interest, and $H$ is some lowpass filter, supported on [0,1]. We will see in Section~\ref{mainsect} that the PIO approximates the counting measure $\tau_I$ as $n\to\infty$, as close as desired. The precise statement will be formulated in Theorem~\ref{maintheo} for a somewhat more general setting.

The computational scheme of the method introduced in this paper, for determining the value of $L$ and computing the coordinates of the points along with their corresponding coefficients, is similar to that  for blind-source signal separation, as introduced in our earlier paper \cite{bspaper}. First, the Hermite moments are computed up to some polynomial degrees to be determined by an input parameter $n$. These moment data are then used as input to the PIO, with output to be thresholded by using an estimate of the minimum of the magnitudes of the coefficients $a_\ell$ as the thresholding parameter. Then the output is separated into $L$ clusters, according to some parameter determined by the minimum separation of the points $\x_\ell$. The location of each of the  $\x_\ell$ is computed by estimation of the maximum value of each of the $L$ cluster. Then using each $\x_\ell$, one by one for  $\ell = 1, \cdots, L$, as input to the PIO yields the value of $a_\ell$. 

The organization of this paper is as follows. The mathematical problem will be formulated precisely in Section~\ref{problemsect} by including the noise/perturbation term and in the general multivariate setting (that is, for any integer $q\ge 1$), and the main result will be stated in Section~\ref{mainsect}. 
In Section~\ref{backsect}, we first recall the notion and some properties of the Hermite functions for the paper to be more self-contained, and then derive a preliminary lemma to facilitate the presentation of our proof of the main result in Section~\ref{pfsect}. 
Application of the mathematical method, developed in this paper, to observatory astronomy and non-invasive diagnosis of caner cells, will be discussed in Section~\ref{application}. 
Here, instead of using the points $\x_{\ell}$ to represent the point-masses, we use suitable groups $X_k$ of these points, so that the shape and color of a point-mass of interest, represented by some $X_k$, can be viewed from the totality of the values $a_{\ell}$, corresponding to the points  $\x_{\ell}$ that constitute the group $X_k$. When this approach is applied to the observation of living (normal and cancer) cells with a super-resolution optical microscopy device, an advantage is that by using the computed results of  $\x_\ell$  as image pixels, we may use the (known) optical resolution of the device as the minimum separation distance among all of the points $\x_{\ell}$. Recall that the minimum separation distance is used to determine the thresholding parameter for finding the number $L$ of clusters in the computational scheme as outlined in the above paragraph. Another advantage is that if the optical resolution is sufficiently high, as compared with the (unknown) size of a (potential) growing cancer cell, represented by $X_k$, being observed, then the size of this cancer cell could be readily estimated by using the number $L_k$ of $\x_{\ell}$ that constitute $X_k$. We end this paper by pointing out, in Section~\ref{remark}, a (mathematical) super-resolution approach, to be introduced and developed in our companion paper \cite{chuigaussian}, where the points $\x_{\ell}$ of the given counting measure are used as centers of a Gaussian sum in order to facilitate the computation of the Hermite moments as input to the PIO, mentioned in the previous paragraph. This approach (studied in \cite{chuigaussian}) also provides a powerful mathematical tool for the separation of real exponential sums, with such important applications as “isotope separation” in nuclear chemistry and “echo cancellation” in signal processing.

\bhag{Problem statement}\label{problemsect}

Let $q\ge 1$ be an arbitrary integer. In this paper, for $\k\in\ZZ_+^q$, $|\k|_1$ denotes the $\ell_1$ norm of $\k$ (that is, the sum of the $q$ components of $\k$). Recall, for example from \cite[Section~5.5]{szego}, that for any multi-integer $\k\in\ZZ^q_+$, the (multivariate) Hermite function $\psi_\k$ is defined via the generating function 
\be\label{hermitegenfunc}
\sum_{\k\in\ZZ^q_+}\frac{\psi_\k(\x)}{\sqrt{2^{|\k|_1} \k!}}\w^\k=\pi^{-1/4}\exp\left(-\frac{1}{2}|\x-\w|^2+|\w|^2/4\right), \qquad \w\in \CC^q.
\ee
As an application of \eref{hermitegenfunc}, it can be deduced that the Hermite functions are invariant under the Fourier transform, up to a multiplicative constant. Precisely, we have
\be\label{hermitefour}
\frac{1}{(2\pi)^{q/2}}\int_{\RR^q}\exp(-i\u\cdot\x)\psi_\k(\u)d\u=(-i)^{|\k|_1}\psi_\k(\x), \qquad \x\in\RR^q.
\ee

Next, for any signed or positive measure $\mu$ with bounded total variation $\|\mu\|_{q,BV}$ on $\RR^q$, the Hermite moments of $\mu$ are defined by
\be\label{hermitemoments}
\hat{\mu}(\k)=\int_{\RR^q}\psi_\k(\x)d\mu(\x), \qquad \k\in\ZZ_+^q.
\ee
If $\mu$ is absolutely continuous with respect to the Lebesgue measure (that is, $d\mu(\u)=F(\u)d\u$), we will abuse the notation by writing $\hat{F}(\k)$ rather than $\hat{\mu}(\k)$. By applying this Hermite-moment notation to both $\tau_I$ and $f_I(\u)d\u$ introduced in the previous section, it follows from \eref{hermitefour} that
\be\label{fourhermitemoment}
\widehat{f_I}(\k)=(-i)^{|\k|_1}(2\pi)^{q/2}\widehat{\tau_I}(\k).
\ee
In other words, Hermite-moment data for $\tau_I$ and $f_I$ are essentially the same.

For application to real-world problems, the counting measure presentation $\tau_I$ of the point-masses in \eref{originalmeasure} is contaminated with additive noise, denoted by  $\tau_c$. In this paper, we will not be concerned with the statistical properties of  $\tau_c$, but assume that it is a (signed or positive) measure with bounded total variation $\|\tau_c\|_{BV}$ on $\RR^q$. In other words, we will replace  \eref{originalmeasure} by the more realistic point-mass representation:
\be\label{perturbedmeasure}
\tau =\sum_{\ell=1}^L a_\ell\delta_{\x_\ell} + \tau_c, 
\ee
with Hermite moments:
\be\label{specialeps}
\hat{\tau}(\k)=\sum_{\ell=1}^L a_\ell\psi_\k(\x_\ell)+\widehat{\tau_c}(\k).
\ee

By further replacing $\widehat{\tau_c}(\k)$ with a somewhat more general perturbation $\epsilon_\k$, the problem to be studied in this paper is to determine the number $L$ of terms in the summation \eref{originalmeasure}, the coordinates of $\x_\ell$ in $\RR^q$, and their corresponding coefficients $a_\ell$, from information of the form 
\be\label{nustardef}
\ns(\k)=\sum_{\ell=1}^L a_\ell\psi_\k(\x_\ell)+\epsilon_\k,
\ee
for only finitely many multi-integers, $\k$.

\bhag{Main results}\label{mainsect}
In this section, we formulate the main theorem of this paper, in finding the number of terms in the summation \eref{nustardef}, and computing $\x_\ell$ and along with their coefficients $a_\ell$, from the data values in \eref{nustardef}, and giving certain error bounds of the approximation errors. Our theorem is similar in spirit to \cite[Theorem~2.4]{bspaper}, with the signal separation operator in \cite{bspaper} replaced by the point-mass isolation operator, to be introduced in Definition~\ref{psodef} below. Throughout this paper, $H:[0,\infty)\to [0,1]$ will denote an infinitely differentiable, non-increasing, even function,   such that $H(t)=1$ for $t\le 1/2$ and $H(t)=0$ for $t\ge 1$. The point-mass isolation operator is defined as follows. 

\begin{definition}\label{psodef}
The family of \textbf{point-mass isolation operators} (PIO) is defined, for any given sequence $\{y_\j\}_{\j\in \ZZ^q_+}$,  by
\be\label{hermitessodef}
\mathcal{T}_n(\{y_\j\}; \x)=n^{-q}\sum_{\j\in\ZZ^q_+}H(\sqrt{|\j|_1}/n)y_\j\psi_\j(\x), \qquad \x\in\RR^q, \ n>0.
\ee
\end{definition}

In the sequel, we will abuse the notation and write $\mathcal{T}_n(\ns;\x)=\mathcal{T}_n(\{\ns(\j)\};\x)$.  For each fixed $n>0$ and $\y\in\RR^q$, when the Hermite functions $\psi_\j(\y)$ are considered as the sequence $y_\j$ in  \eref{psodef},  we have a kernel function $\Phi_n(\x,\y)$ of two variables $\x, \y$, associated with the PIO $\mathcal{T}_n$, namely: 
\be\label{hermite_kernel}
\Phi_n(\x,\y):=\mathcal{T}_n(\{\psi_\j(\y)\}; \x)=n^{-q}\sum_{\j\in\ZZ^q_+} H\left(\frac{\sqrt{|\j|_1}}{n}\right)\psi_\j(\x)\psi_\j(\y), \qquad n>0, \x,\y\in\RR^q.
\ee
We will list various properties of this kernel function in Lemma~\ref{kernlemma}. In particular, in Lemma~\ref{kernlemma}(a), we will see that the family of $\Phi_n, n>0,$ constitutes an approximate identity, so that when the perturbations $\epsilon_\k$ in \eref{nustardef} are small, we have
$$\mathcal{T}_n(\ns;\x) \approx \sum_{\ell=1}^L a_\ell\delta_{\x_\ell}(\x),$$ 
in the sense to be made precise in our main theorem Theorem~\ref{maintheo}, to be stated below.

But first, let us point out that
\be\label{taunnustar}
\mathcal{T}_n(\ns;\x)=\sum_{\ell=1}^L a_\ell\Phi_n(\x,\x_\ell) +\mathcal{T}_n(\{\epsilon_\j\};\x).
\ee
To provide a general mathematical tool beyond the objectives of this paper, we will replace $\mathcal{T}_n(\{\epsilon_\j\};\x)$ in \eref{taunnustar} by some function $E_n:\RR^q\to\RR$. In other words, we formulate our main theorem in terms of a sum of the form
\be\label{generaltaun}
\mathbb{T}_n(\x)=\sum_{\ell=1}^L a_\ell\Phi_n(\x,\x_\ell)+E_n(\x).
\ee

To facilitate the statement of the theorem, we introduce the following notations.
\be\label{Metcdef}
M=\sum_{_\ell=1}^L|a_k|,\ \mu=\min_{1\le _\ell \le L}|a_\ell|,\ \eta=\min_{1\le k\not=j\le L}|\x_\ell-\x_j|, \ B=\max_{1\le \ell\le L}|\x_\ell|_\infty.
\ee
Analogous to \cite[Theorem~2.4]{bspaper}, our main theorem is stated as follows. In this statement, we will engage three positive constants, namely:  $A_2$ and $\alpha$, to be introduced in Lemma~\ref{kernlemma} of Section~\ref{backsect}; and $\gamma$, to be defined in \eref{gammadef} in Section~\ref{pfsect} in terms of the quantities in \eref{Metcdef} . 

\begin{theorem}\label{maintheo}
For $n\ge 1$, let
\be\label{largeset}
\mathcal{G}_n =\{\x\in \RR^q : |\mathbb{T}_n(\x)|\ge A_2\mu/2\},
\ee
and assume that
\be\label{noisecond}
|E_n(\x)|\le A_2\mu/8, \qquad \x\in\RR^q.
\ee
Then for all sufficiently large values of $n$, the following statements hold.
\begin{enumerate}
\item[{\rm (a)}]
 There exists a partition $\mathcal{G}_{n,\ell}$, $\ell=1,\ldots, L$,  of $\mathcal{G}_n$ such that
\begin{enumerate}
\item[(i)] for $\ell=1,\ldots, L$, $\x_\ell\in \mathcal{G}_{n,\ell}$,
\item[(ii)] for $\ell=1,\ldots, L$, $\mathsf{diam}( \mathcal{G}_{n,\ell})\le 2\gamma/n\le \eta/2$, and
\item[(iii)] for $\ell, j=1,\ldots, L$, $\ell\not=j$, $\mathsf{dist} (\mathcal{G}_{n,\ell},\mathcal{G}_{n,j})\ge\eta/2$.
\end{enumerate}
\item[{\rm (b)}] Let 
\be\label{capNcond}
N\ge \max(1,\frac{2\gamma}{\a})n
\ee
and for $\ell=1,\cdots,L$, set
\be\label{estcenter}
\widehat{\x}_{N,n,\ell}=\arg\max_{\x\in \mathcal{G}_{N,\ell}}|\mathbb{T}_n(\x)|.
\ee
Then
\be\label{centeresterror}
|\widehat{\x}_{N,n,\ell}- \x_\ell| \le 2\gamma/N \le \a/n,
\ee
and for some suitably small positive $\epsilon$ (cf. \eref{epssmall}),
\be\label{basicampfound}
|\mathbb{T}_n(\widehat{\x}_{N,n,\ell})-a_\ell\Phi_n(\widehat{\x}_{N,n,\ell},\widehat{\x}_{N,n,\ell})| \le 5\epsilon.
\ee 
\end{enumerate} 
\end{theorem}

An  example of importance in this paper is when the observed Hermite moments are contaminated with some additive perturbed measure in \eref{perturbedmeasure}.  In this case, we will consider, for $\k\in\ZZ^q_+$,
\be\label{specialepsbis}
\epsilon_\k=\widehat{\tau_c}(\k)=\int_{\RR^q}\psi_\k(\u)d\tau_c(\u)
\ee
and show in Lemma~\ref{kernlemma} that there exists a positive constant $A_1$, that depends only on $H$ and $q$, such that
$$
|\Phi_n(\x,\y)| \le A_1, \qquad \x,\y\in \RR^q.
$$
Hence, for $\epsilon_\k$ as given in \eref{specialepsbis}, it follows that
\be\label{enest}
|E_n(\x)| =\left|n^{-q}\sum_{\j\in\ZZ^q_+}H(\sqrt{|\j|_1}/n)\epsilon_\j\psi_\j(\x)\right|=\left|\int_{\RR^q} \Phi_n(\x,\y)d\tau_c(\y)\right| \le A_1\|\tau _c\|_{q,BV}.
\ee
Consequently, the condition \eref{noisecond} stipulates that the measure $\tau_c$ must be assumed not to dominate any of the terms $a_\ell\delta_{\x_\ell}$ in \eref{originalmeasure}. If the perturbation would dominate at least one of the point measures, then it is not reasonable to expect a solution of the problem.

\bhag{Background}\label{backsect}
In this section, we recall some properties of the Hermite functions and derive some preliminary technical results. In Sub-Section~\ref{hermitesect}, we review the properties of  Hermite functions; and in Sub-Section~\ref{prelimressect}, we derive the auxiliary results needed in the proof of Theorem~\ref{maintheo}. Throughout the rest of this paper, we follow the following convention regarding constants.

\noindent\textbf{Constant convention:}\\
\textit{ The symbols $c, c_1, c_2, \cdots$ will denote generic positive constants depending only upon $H$, $q$, and the number $S$ introduced below.
 Their values may differ at different occurrences, even within a single formula.
 On the other hand, constants denoted by capital letters will retain their values.}

\subsection{Hermite functions}\label{hermitesect}
The univariate Hermite functions $\{\psi_j\}$ satisfy the Rodrigues' formula; i.e., \cite[Formula~(5.5.3)]{szego}
\be\label{hermitedef}
\psi_j(x)= \frac{(-1)^j}{\pi^{1/4}2^{j/2}\sqrt{j!}}\exp(x^2/2)\left(\frac{d}{dx}\right)^j (\exp(-x^2)), \qquad x\in\RR, \ j=0,1,\cdots,
\ee
and satisfy the orthogonality relation for $j, k=0,1,\cdots$ (\cite[Formula~(5.5.1)]{szego}):
\be\label{uniortho}
\int_\RR \psi_j(x)\psi_k(x)dx=\left\{\begin{array}{ll}
1, &\mbox{ if $j=k$,}\\
0, &\mbox{otherwise}.
\end{array}\right.
\ee
Furthermore, for all $j=0,1,2,\cdots$ and $x\in\RR$, the Hermite functions satisfy the differential equation (\cite[Formula~(5.5.2)]{szego}):
\be\label{diffeqn}
\psi_j''(x)=(x^2-2j-1)\psi_j(x),
\ee
and the three-term recurence relation (cf. \cite[Formula~(5.5.8)]{szego})
\bea\label{recurrence}
x\psi_{j-1}(x)&=&\sqrt{\frac{j}{2}}\psi_j(x) + \sqrt{\frac{j-1}{2}}\psi_{j-2}(x),\quad j=2,3,\cdots,\nonumber\\
&&\psi_0(x)=\pi^{-1/4},\ \psi_1(x)=\sqrt{2}\pi^{-1/4}x\exp(-x^2/2).
\eea
 
In the sequel, for any $\lambda>0$,  let $\mathbb{P}_\lambda$ denote the set of all polynomials of degree $<\lambda^2$, and $\Pi_\lambda$ denote the linear span of $\psi_j$, $0\le j<\lambda^2$, Thus, an element of $\Pi_\lambda$ has the form $P(x)\exp(-x^2/2)$, where $P\in\mathbb{P}_\lambda$.
We have the Bernstein inequality (cf. \cite{freud1972direct}):
\be\label{unibern}
\|P'\|_{p,\RR} \le cn\|P\|_{p,\RR}, \qquad P\in \Pi_n, \ 1\le p\le \infty.
\ee
Next, we note the following identities for the reproducing kernel. By applying \eref{recurrence} and following  the proof of (cf. \cite[Theorem~3.2.2]{szego}), it can be deduced that
\be\label{christdarboux}
\sum_{j=0}^{n-1} \psi_j(x)\psi_j(y)=\sqrt{\frac{n}{2}}\frac{\psi_n(x)\psi_{n-1}(y)-\psi_n(y)\psi_{n-1}(x)}{x-y},
\ee
and hence, passing to the limit, we have
\be\label{christondiag}
K_n(x)=\sum_{j=0}^{n-1} \psi_j(x)^2=\sqrt{\frac{n}{2}}\{\psi_n'(x)\psi_{n-1}(x)-\psi_n(x)\psi_{n-1}'(x)\}.
\ee
In view of  \eref{diffeqn}, it is easy to verify that
\be\label{christder}
K_n'(x)=\sqrt{\frac{n}{2}}\{\psi_n''(x)\psi_{n-1}(x)-\psi_n(x)\psi_{n-1}''(x)\}=-\sqrt{2n}\psi_n(x)\psi_{n-1}(x).
\ee

In addition, we recall the upper bounds
\be\label{hermitebds}
|\psi_n(x)|\le\left\{\begin{array}{ll}
cn^{-1/4}, &\mbox{ if $|x|\le n^{1/2}(1-n^{-2/3})$,}\\
cn^{-1/12}, &\mbox{if $x\in\RR$.}
\end{array}\right.
\ee
for the Hermite functions derived in \cite{askey1965mean} and 
\be\label{unichristbd}
|K_n(x)|\le cn, \qquad x\in\RR, \qquad |K_n(x)| \ge cn, \quad |x|\le c_1 n
\ee
for the kernels, proved in \cite[Theorem~3.2.5]{mhasbk}.
Furthermore, by applying \eref{christder} and \eref{hermitebds}, we obtain  
\be\label{christcont}
|K_n(x)-K_n(y)|\le c|x-y|\left\{\begin{array}{ll}
1, &\mbox{ if $x, y\in c_1\sqrt{n}$,}\\
n^{1/3}, &\mbox{ if $x,y\in\RR$.}
\end{array}\right.
\ee
We also recall the Mehler formula \cite[Formula~(6.1.13)]{andrews_askey_roy} which states that
\be\label{mehler}
\sum_{j=0}^\infty \psi_j(y)\psi_j(z)r^j= \frac{1}{\sqrt{\pi (1-r^2)}}\exp\left(\frac{2yzr-(y^2+z^2)r^2}{1-r^2}\right)\exp(-(y^2+z^2)/2),  
\ee
for all $y, z\in\RR$ and $|r|<1$. 

Let us now turn to the multivariate setting. The multivariate Hermite functions
\be\label{multihermitedef}
\psi_\k(\x)=\prod_{j=1}^q \psi_{k_j}(x_j),
\ee
defined by using tensor-products, with $\k= (k_1, \cdots, k_q)$, also satisfy the following orthonormality condition and the Mehler formula: 

\be\label{hermiteortho}
\int_{\RR^q} \psi_\j(\z)\psi_{\bs\ell}(\z)d\z =\delta_{\j,\bs\ell},\qquad \j,\bs\ell\in \ZZ^q_+
\ee
and
\be\label{multimehler}
\sum_{\j\in\ZZ^q_+} \psi_\j(\y)\psi_\j(\z)r^{|\j|_1}= \frac{1}{(\pi (1-r^2))^{q/2}}\exp\left(\frac{2\y\cdot\z r-(|\y|^2+|\z|^2)r^2}{1-r^2}\right)\exp(-(|\y|^2+|\z|^2)/2), 
\ee
for all  $\y, \z\in\RR^q$ and $ |r|<1$.

Finally, by applying \eref{unichristbd}, it is not difficult to show that
\be\label{multichristlowbd}
\sum_{\sqrt{|\j|_1}< n} |\psi_\j(\x)|^2  \ge \sum_{\sqrt{|\j|_\infty}< n/\sqrt{q}} |\psi_\j(\x)|^2=\prod_{\ell=1}^q\sum_{0\le j_\ell< n^2/q}|\psi_{j_\ell}(x_\ell)|^2 \ge cn^q, 
\ee
for all $\x$ with $|\x|_\infty \le cn$, where the sup-norm notation is used here and later in this paper.

\subsection{Preliminary results}\label{prelimressect}

In the following, $\mathbb{P}_\lambda^q$  denotes the class of all polynomials in $q$ variables with total  degree $<\lambda^2$; and $\Pi_\lambda^q$  denotes the class of all functions of the form $\x\mapsto P(\x)\exp(-|\x|^2/2)$, $P\in \mathbb{P}_\lambda^q$. Let us recall the lowpass filter $H$ defined in Section~\ref{mainsect} and the kernel $\Phi_n$ introduced in \eref{hermite_kernel}.  Also note that the kernel $\Phi_n\in \Pi_n^q$  is a function of both $\x$ and $\y$. In the following, we summarize some properties of this kernel.
\begin{lemma}\label{kernlemma} 
Let $S>q$ be any given integer. There exist constants $A_1, A_2, C, C_1>0$, such that each of the following statements hold.
\begin{enumerate}
\item[{\rm (a)}] For $\x,\y\in\RR^q$, $n=1,2,\cdots$,
\be\label{hermite_localization}
|\Phi_n(\x,\y)| \le \frac{A_1}{\max(1,(n|\x-\y|)^S)}.
\ee
\item[{\rm (b)}] 
%For $\x\in\RR^q$, $n=1,2,\cdots$,
%\be\label{phinintest}
%\int_{\RR^q}|\Phi_n(\x,\y)|d\y \le cn^{-q}.
%\ee
%\end{enumerate}
%
%\begin{enumerate}
%\item[{\rm (c)}] 
For  $n=1,2,\cdots$,
\be\label{philowbd}
|\Phi_n(\x,\x)|\ge A_2, \qquad |\x|_\infty\le Cn.
\ee
\item[{\rm (c)}] For $n\ge 1$, $|\x|_\infty, |\y|_\infty\le Cn$, 
\be\label{summdiagstable}
|\Phi_n(\x,\x)-\Phi_n(\y,\y)|\le C_1n^{-q}|\x-\y|.
\ee
\item[{\rm (d)}]   There exists $\a>0$ such that
\be\label{schwarzcond}
0\le \Phi_n(\x,\y)\le \Phi_n(\y,\y), \qquad \x, \y\in\RR^q, \ |\x-\y|\le \a/n,\   |\y|_\infty\le Cn, \ n\ge 1.
\ee
\end{enumerate}
\end{lemma}
 
Lemma~\ref{kernlemma}(a) plays a critical role in our proof of Theorem~\ref{maintheo}. Estimates analogous to \eref{hermite_localization} were studied by many authors, e.g., \cite{dziubanski1997triebel, epperson1997hermite}, typically using fairly complicated proofs. Our proof is much simpler, and follows directly from the Mehler formula and the   following  result, proved in \cite[Theorem~4.3]{tauberian} also using elementary techniques.
 
\begin{theorem}\label{maintaubertheo}
Let $\mu$ be an extended complex-valued measure on $[0,\infty)$ with $\mu(\{0\})=0$, and assume that there exist $Q, r>0$, such that the following conditions are satisfied.
\begin{enumerate}
\item 
\be\label{muchristbd}
\tn\mu\tn_Q:=\sup_{u\in [0,\infty)}\frac{|\mu|([0,u))}{(u+2)^Q} <\infty.
\ee
\item There are constants $c, C >0$,  such that
\be\label{muheatgaussbd}
\left|\int_\RR \exp(-u^2t)d\mu(u)\right|\le c_1t^{-C}\exp(-r^2/t)\tn\mu\tn_Q, \qquad 0<t\le 1.
\ee 
\end{enumerate}
Furthermore, let $H:[0,\infty)\to\RR$, $S>Q+1$ be an integer, and that there exists a measure $H^{[S]}$ such that
\be\label{Hbvcondnew}
H(u)=\int_0^\infty (v^2-u^2)_+^{S}dH^{[S]}(v), \qquad u\in\RR,
\ee
and
\be\label{Hbvintbdnew}
V_{Q,S}(H)=\max\left(\int_0^\infty (v+2)^Qv^{2S}d|H^{[S]}|(v), \int_0^\infty (v+2)^Qv^Sd|H^{[S]}|(v)\right)<\infty.
\ee
Then for all integers $n\ge 1$,
\be\label{genlockernest}
\left|\int_0^\infty H(u/n)d\mu(u)\right| \le c\frac{n^Q}{\max(1, (nr)^S)}V_{Q,S}(H)\tn\mu\tn_Q.
\ee
\end{theorem}
 
\noindent
\textsc{Proof of Lemma~\ref{kernlemma}.}\\

In this proof, we define
$$
\mu(u)=\mu_{\x,\y}(u)= \sum_{ \sqrt{|\j|_1}<u} \psi_\j(x)\psi_\j(\y).
$$
Then
\be\label{sumasint}
\sum_{\j\in\ZZ^q_+} \psi_\j(\x)\psi_\j(\y)e^{-t|\j|_1}=\int_0^\infty e^{-tu^2}d\mu_{\x,\y}(u),\qquad \Phi_n(\x,\y)=n^{-q}\int_0^\infty H\left(\frac{u}{n}\right)d\mu_{\x,\y}(u).
\ee
Writing $\x, \y$ in place of $\y, \z$, $e^{-t}$ in place of $r$, and completing the squares, we may deduce from \eref{multimehler} that
\be\label{hermite_heatkern}
\sum_{\j\in\ZZ^q_+} \psi_\j(\x)\psi_\j(\y)e^{-t|\j|_1}=\frac{e^{qt/2}}{(2\pi\sinh t)^{q/2}}\exp\left(-\frac{2}{\sinh t}|\x-\y|^2\right)\exp(-(1/2)\tanh (t/2)(|\x|^2+|\y|^2)),
\ee
from which it follows that, for $0<t<1$,
\be\label{hermite_gaussbd}
\left|\sum_{\j\in\ZZ^q_+} \psi_\j(\x)\psi_\j(\y)e^{-t|\j|_1}\right|\le c_1t^{-q/2}\exp\left(-c_2\frac{|\x-\y|^2}{t}\right).
\ee
Thus, we may conclude that \eref{muheatgaussbd} holds. Moreover, according to \cite[Proposition~4.1]{frankbern}, it also follows from \eref{hermite_gaussbd}  that
\be\label{multichristbd}
\sum_{\sqrt{|\j|_1}< u} |\psi_\j(\x)|^2 \le cu^q, \qquad u\ge 1,
\ee
from which, we obtain \eref{muchristbd} with $Q=q$. Now, since $H$ is an even and infinitely differentiable function, supported on $[-1,1]$, the conditions \eref{Hbvcondnew} and \eref{Hbvintbdnew} are satisfied. Therefore, Theorem~\ref{maintaubertheo} implies that \eref{hermite_localization} holds. 

%To prove \eref{phinintest}, we  observe that
%$$
%\int_{\RR^q}|\Phi_n(\x,\y)|d\y \le c\left\{\int_{\y : |\x-\y|\le 1/n} d\y +\frac{1}{n^S}\int_{\y: |\x-\y|\ge 1/n}\frac{d\y}{|\x-\y|^S}\right\}\le cn^{-q}.
%$$

  Next, integration by parts in \eref{sumasint}, followed by setting $u$ in place of $u/n$, yields
$$
n^q\Phi_n(\x,\x)=-\int_0^\infty H'(u)\mu_{\x,\x}(nu)du.
$$
Hence, since $H$ is a lowpass function, we have
\be\label{pf1eqn1}
n^q\Phi_n(\x,\x)=-\int_{1/2}^1 H'(u)\mu_{\x,\x}(nu)du = -\int_{1/2}^1 H'(u)\left\{\sum_{\sqrt{|\j|_1}<nu}\psi_\j(\x)^2\right\}du.
\ee
In addition, since $H'(u)\le 0$ for all $u\in [1/2,1]$ and $H'\not\equiv 0$, there is a sub-interval $[a,b]\subseteq [1/2,1]$ such that $-H'(u)\ge c$ for $u\in [a,b]$. Hence, \eref{multichristlowbd} leads to \eref{philowbd}.
 
To prove part (c), we apply \eref{christcont} one coordinate of $\x, \y$ at a time, to deduce 
$$
\left|\sum_{\sqrt{|\j|_1}<nu}\psi_\j(\x)^2-\sum_{\sqrt{|\j|_1}<nu}\psi_\j(\y)^2\right| \le c|\x-\y|, \qquad 1/2\le u\le 1, \ |\x|_\infty,|\y|_\infty \le cn.
$$
Consequently, \eref{pf1eqn1} leads to \eref{summdiagstable}. Finally, in view of \eref{philowbd} and \eref{hermite_localization}, we see that
$$
\max_{\z\in\RR^q}|\Phi_n(\z,\y)|\le c\le c_1\Phi_n(\y,\y), \qquad |\y|_\infty\le c_2n,
$$
from which we may deduce, by applying \eref{unibern}, that for $|\y|_\infty\le cn$, 
$$
|\Phi_n(\x,\y)-\Phi_n(\y,\y)|\le cn|\x-\y|\max_{\z\in\RR^q}|\Phi_n(\z,\y)|\le cn|\x-\y|\Phi_n(\y,\y).
$$
This, together with \eref{philowbd}, leads to the completion of the proof of part (d) in the statement of the lemma.
\qed

%%%%%%

\bhag{Proof of Theorem~\ref{maintheo}.}\label{pfsect}

\noindent\textsc{Proof of Theorem~\ref{maintheo}}(a). Let
\be\label{gammadef}
\gamma=\max\left(1, \left(\frac{8A_1M}{A_2\mu}\right)^{1/S}\right)
\ee
and
\be\label{partdef}
I_\ell=\{\x\in\RR^q : |\x-\x_\ell|\le \gamma/n\}, \qquad \ell=1,\cdots, L.
\ee
Then for all $n$ that satisfy
\be\label{lambdalowbd1}
n\ge \max(1, 4\gamma/\eta, B/C),
\ee
where $C$ is the constant that appears in \eref{schwarzcond}, the estimate \eref{lambdalowbd1} implies that
\be\label{pf2eqn1}
\mathsf{diam}(I_\ell) \le 2\gamma/n\le \eta/2,\quad \mathsf{dist}(I_\ell,I_j)\ge\eta/2, \qquad \ell, j=1,\cdots, L,\ \ell\not=j.
\ee
On the other hand, by applying \eref{hermite_localization} and recalling \eref{noisecond} and \eref{gammadef}, it is easy to deduce that for $\x\in \RR^q\setminus \bigcup_{\ell=1}^L I_\ell$, 
\be\label{pf2eqn2}
|\mathbb{T}_n(\x)| \le \sum_{\ell=1}^L |a_\ell|\frac{A_1 }{\max(1, (n
|\x-\x_\ell|)^S)}+|E_n(\x)| \le \frac{A_1M }{\gamma^S}+A_2\mu/8\le A_2\mu/4;
\ee
so that for $\x\in \mathcal{G}_{n}$, we have $\x\in I_\ell$ for some $\ell$. In view of \eref{pf2eqn1}, this $\ell$ is unique. Thus it follows that the collection  $\{\mathcal{G}_{n,\ell}\}_{\ell=1}^L$, where $\mathcal{G}_{n,\ell} :=I_\ell\cap \mathcal{G}_n$, constitutes a partition of  $\mathcal{G}_n$ that satisfies the properties (ii) and (iii) of Theorem~\ref{maintheo}(a).

The same argument as in the derivation of \eref{pf2eqn2} also shows that 
\be\label{pf2eqn3}
|\mathbb{T}_n(\x_\ell)-a_\ell\Phi_n(\x_\ell,\x_\ell)| \le A_2\mu/4.
\ee
Consequently, the result in \eref{philowbd} asserts that
\be\label{pf2eqn4}
|\mathbb{T}_n(\x_\ell)| \ge A_2\mu-A_2\mu/4\ge A_2\mu/2.
\ee
Therefore, we may conclude that  $\x_\ell\in \mathcal{G}_{n,\ell}$, for every $\ell = 1, \cdots, L$, as stated in item (i) of Theorem~\ref{maintheo}(a).
\qed

\noindent\textsc{Proof of Theorem~\ref{maintheo}}(b). In this proof, we assume that 
\be\label{largelambda}
n\ge \max\left(1, 4\gamma/\eta, 2B/C, 4\gamma/\sqrt{C}, \left(\frac{A_2}{4C_1\gamma}\right)^{1/(q+1)}, \frac{2}{\eta}\left(\frac{16A_1M}{A_2\mu}\right)^{1/S}, \left(\frac{32C_1\gamma M}{A_2\mu}\right)^{1/(q+1)}\right),
\ee
where $C$ is the constant introduced in  \eref{philowbd} and $C_1$ is the constant introduced in \eref{summdiagstable}.
Then in view of \eref{noisecond}, we have
\be\label{epssmall}
\epsilon :=\max(|E_n(\x_\ell)|, |E(\widehat{\x}_{N,n,\ell})|)+\frac{2^SA_1M }{(n\eta)^S}+\frac{2MC_1\gamma}{ n^{q+1}}\le \frac{\mu A_2}{4}.
\ee
Also, observe that the estimate \eref{centeresterror} follows from the definition and Theorem~\ref{maintheo}(a)(i).
We continue to use the same notation as in the proof of Theorem~\ref{maintheo}(a), and 
write $\widehat{\x}_\ell=\widehat{\x}_{N,n, \ell}$. Since $\widehat{\x}_\ell\in I_\ell$ and each $\x_j\in I_j$, we have  $|\widehat{\x}_\ell-\x_j|\ge \eta/2$ for $j\not=\ell$, by applying  \eref{pf2eqn1}. Therefore, using \eref{hermite_localization}, we may conclude, as in the derivation of \eref{pf2eqn2}, that
\be\label{pf4eqn1}
|\mathbb{T}_n(\widehat{\x}_\ell)-a_\ell\Phi_n(\widehat{\x}_\ell,\x_\ell)| \le |E_n(\widehat{\x}_\ell)|+\frac{2^SA_1M }{(n\eta)^S},
\ee
and
\be\label{pf4eqn2}
|\mathbb{T}_n(\x_\ell)-a_\ell\Phi_n(\x_\ell,\x_\ell)| \le |E_n(\x_\ell)|+\frac{2^SA_1M }{(n\eta)^S}.
\ee
In the following, for convenience and for this proof only, set
$$
\delta :=\max(|E_n(\x_\ell)|, |E(\widehat{\x}_\ell)|)+\frac{2^SA_1M }{(n\eta)^S}.
$$
Then \eref{pf4eqn1} and  \eref{schwarzcond} together imply that
\be\label{pf4eqn3}
|\mathbb{T}_n(\widehat{\x}_\ell)|\le |a_\ell||\Phi_n(\widehat{\x}_\ell,\x_\ell)|+\delta \le |a_\ell|\Phi_n(\x_\ell,\x_\ell)+\delta.
\ee
Since $\x_\ell\in\mathcal{G}_{n,\ell}$, the definition \eref{estcenter} of $\widehat{\x}_\ell$ together with \eref{pf4eqn2} implies that
\be\label{pf4eqn4}
|\mathbb{T}_n(\widehat{\x}_\ell)|\ge |\mathbb{T}_n(\x_\ell)|
\ge |a_\ell|\Phi_n(\x_\ell,\x_\ell)-\delta.
\ee
In view of \eref{pf4eqn3} and \eref{pf4eqn4}, we obtain
\be\label{pf4eqn5}
\left||\mathbb{T}_n(\widehat{\x}_\ell)|-|a_\ell|\Phi_n(\x_\ell,\x_\ell)\right| \le \delta.
\ee
Since $n\ge 4\gamma/\sqrt{C}$, it follows from \eref{centeresterror} that 
$ |\widehat{\x}_\ell|_\infty \le Cn$. Therefore, Lemma~\ref{kernlemma}(c) yields
$$
|\Phi_n(\widehat{\x}_\ell,\widehat{\x}_\ell)-\Phi_n(\x_\ell,\x_\ell)|\le 2C_1\gamma n^{-q-1}.
$$
Thus, \eref{pf4eqn5} leads to
$$
\left||\mathbb{T}_n(\widehat{\x}_\ell)|-|a_\ell|\Phi_n(\widehat{\x}_\ell,\widehat{\x}_\ell)\right| \le \delta +2MC_1\gamma n^{-q-1}=\epsilon.
$$

In the remainder of this proof, let $\mathbb{T}_n(\widehat{\x}_\ell)=: Re^{i\phi}$, $a_\ell\Phi_n(\widehat{\x}_\ell,\widehat{\x}_\ell)=: re^{i\theta}$, $\rho :=\Phi_n(\widehat{\x}_\ell,\x_\ell)/\Phi_n(\widehat{\x}_\ell,\widehat{\x}_\ell)$, and consider $\psi=\theta-\phi$. 
Since $|\widehat{\x}_\ell-\x_\ell|\le 2\gamma/N\le \a/n$, our assumptions imply that $\rho\ge 0$. Further, \eref{pf4eqn1} and \eref{pf4eqn5} can be re-written in the form of
\be\label{pf4eqn6}
|R-r\rho e^{i\psi}|\le \epsilon, \quad |R-r| \le \epsilon.
\ee
Hence, we obtain
\be\label{pf4eqn7}
|1-\rho e^{i\psi}| \le \frac{2\epsilon}{r}; \quad \mbox{ i.e., } (1-\rho)^2+4\rho\sin^2(\psi/2)\le \frac{4\epsilon^2}{r^2}.
\ee
Observe that $r\ge \mu A_2\ge 4\epsilon$,  \eref{pf4eqn7} leads to $\rho\ge 1-2\epsilon/r\ge 1/2$, and
$$
|1-e^{i\psi}|=2|\sin(\psi/2)| \le \frac{2\epsilon}{r-2\epsilon}\le 4\epsilon.
$$
Therefore, in view of \eref{pf4eqn6}, we may now conclude that
$$
|R-r e^{i\psi}| \le |R-r| +r|1-e^{i\psi}| \le 5\epsilon.
$$
This completes the proof of \eref{basicampfound}. \qed

\bhag{Application to observational astronomy and noninvasive health care}\label{application}

In the problem statement of this paper, with the (perturbed) counting measure in  \eref{perturbedmeasure} as the source data, the points $\x_{\ell}$,  $\ell = 1, \cdots, L$, represent only the locations of the point-masses in $\RR^q$, but do not convey any information concerning their sizes, shapes, etc. However, for most real-world applications, point-masses of interest are usually identified by their geometric shapes as well as color shades or grayscale intensities. For this purpose, a point-mass should be considered as a group of (more than one) $\x_{\ell}$, by partitioning the set of all $\{\x_{\ell}\}$ into $N < L$ subsets $X_k$ as follows. For each $k = 1, \cdots, N$, let $L_k$ be the cardinality of $X_k$ and we re-label the points $\x_{\ell}\in X_k$  by $\x_{j,k}$, $j=1,\cdots, L_k$. Then we introduce the (disjoint) groups of points by
\be\label{unionexp}
X_k:=\{\x_{j,k} : j=1,\cdots, L_k\}, \qquad k=1,\cdots, N.
\ee   
  Note that $L_1+\cdots+L_N= L$. This partition is equivalent to the partition of the source data $\tau$ in \eref{perturbedmeasure}, namely: 
\be\label{taupartitiondef}
\tau=\sum_{k=1}^N\tau_k + \tau_c,
\ee
where
\be\label{taukdef}
\tau_k:=\sum_{j=1}^{L_k} a_{j,k}\delta_{\x_{j,k}}.
\ee
We will consider a totality of $N$ ``point-masses”, represented by the $X_k$ where $k=1,\cdots, N$, as defined in \eref{unionexp}. To unify our discussion, we do not exclude the possibility of a single point $\x_{\ell}$ as one of the $X_k$. But if the index set $\mathcal{I}_k=\{1,\cdots, L_k\}$ is sufficiently large, this newly formed “point-mass” representation $X_k$ should exhibit some geometric shape, which is “visible” from the attributes of the points $\x_{j,k}$ that constitutes $X_k$, namely via the template:
\be\label{templatedef}
A_k=[a_{j,k}]_{j\in \mathcal{I}_k},
\ee       
where $(j,k)$ indicates the position $\x_{j,k}$ in $\RR^q$. 

To apply this mathematical formulation to the two problems briefly mentioned in the abstract of this paper, let us first consider each group $X_k$ of points in \eref{unionexp} as a galaxy in the universe, with its stars represented by the points $\x_{j,k}$ where $j\in \mathcal{I}_k$. As is well-known, galaxies are identified by their shapes and colors. Indeed, the most widely used classification scheme, proposed by Edwin Hubble (in whose honor, the Hubble space telescope, HST, was named) consists of spiral galaxies, elliptical galaxies, and irregular galaxies. Being the most common type, a spiral galaxy, such as our own Milky Way, is a rotating disk of stars and nebulae, surrounded by a shell of dark matter and with a bright central region at the core, called “galactic bulge”. On the other hand, the shape of an elliptic galaxy is ellipsoidal or ovoid, with size ranging from only a few thousand light-years to over hundreds of thousand light-years in diameter. Irregular galaxies have no particular shape. Full of gas and dust, most irregular galaxies are very bright. Those that are over 13 billion light-years away (implying that these galaxies are very young as we see them, and most probably have lots of star formation going on within them) are mainly irregular. As to color classification, many young stars exist in “blue” regions, since such stars live fast and die young, consuming fuel at a high rate to maintain high temperatures that emit “blueish” hot radiation. On the other hand, old stars exist in “red” regions, since such stars have swollen and cooled, and emit “reddish” radiation. In addition, when free protons capture free electrons in a cloud of ionized hydrogen, called an H-II (or H-two) region, light of various wavelengths, in “red/pink” color, is emitted, as electrons hop down through energy levels. H-II regions are ionized by ultraviolet radiation from hot stars, indicating the birth of new stars. 

It is interesting to point out that the “Abell 2744 Y1” galaxy, discovered by the HST in July 2014, is some 13 billion light-years away! Assuming that the Big Bang occurred 13.7 billion years ago and that it was the very beginning of the universe, this galaxy, as observed from our solar system, was at most 700 million years old.  So, it would be most fascinating if it would be possible to discover galaxies that are even farther away. Unfortunately, the Hubble telescope does not have the capability of capturing images beyond the 1,700 nm wavelength in the near-infrared spectrum; and although the Spitzer is an infrared space telescope, it has much lower resolution than Hubble and is limited to capturing images up to 11.5 billion light-years away in space. The great news is that after many years of preparation, the James Webb space telescope (JWST or Webb), as successor of both Hubble and Spitzer, is scheduled to launch in October, 2018, and to orbit around the Sun, while staying in line with the Earth and keeping a distance of about 1 million miles away from us.  Different from Hubble and similar to Spitzer, JWST is strictly an infrared space telescope, but is far more powerful than both. With the size of the primary mirror  2.7 times in diameter larger than that of Hubble and operating in much colder deep space than Spitzer, the Webb has the capability of capturing high-resolution images in mid-infrared wavelengths, and is expected to see stars that are 13.5 billion light-years away! One of the missions of JWST is to search the light of the first stars and galaxies after Big Bang, with the objective in helping scientists to study the evolution of galaxies, stars and planets.

Going from the very large to the very small, we next discuss the application of the above mathematical formulation to observing the growth of cancer cells among living cells. Again, instead of considering the points $\x_{\ell}$ as cells, we use each group $X_k$ of points as introduced in \eref{unionexp} to represent a cell, for the purpose of observing shapes, size, and color shades.  More precisely, for a digital image of any cell represented by $X_k$, the points $\x_{\ell}$ (that constitute the group $X_k$) and the corresponding coefficients $a_{\ell}$ of the counting measure \eref{taukdef} can be used to represent the pixel locations and pixel values of this cell digital image. The advantage of this approach is that the size of the cell can be estimated by using the number $L_k$ of $\x_{j,k}\in X_k$ and that the corresponding template $A_k$ defined in \eref{templatedef} reveals the shape and colors of the cell, assuming that the resolution of the cell digital image is sufficiently high.  In this regard, we note that size of human cells differs, ranging from 3,850 nm to 120,000 nm in diameter, with the male sperm among the smallest and female egg the largest. About 70\% of all human cells are red blood cells, which are of donut shape (without holes) and with size ranging from around 6,200 nm  and 8,200 nm in disk diameter. It is important to point out that cancer cells and normal cells have different sizes, shapes, and colors, particularly in the nucleus, called the “brain” of the cell. For example, under the light microscope, cancer cells often exhibit much more variability in size, with some larger and some smaller than normal cells, their shapes are irregular, being fractal-like, and the nucleus of a cancer cell is both larger and darker (in color) than a normal cell. Furthermore, since cancer cells grow and spread fast, it is of utmost importance to be able to observe the rate of their growth before they spread to other organs of the human body.

Currently, the most common diagnosis procedure of cancer cells is to perform biopsies and to stain the tissue samples with certain antibodies and biomarkers, and then to study the specific markers under a light microscope. This procedure is not only invasive, but also not very accurate, with only 85\% success rate. Furthermore, it is even more invasive to perform follow-up biopsies for monitoring cancer growth. It is clear that the much more sensible way is to rely on digital imaging, as proposed above. However, there are at least two obstacles to overcome. Firstly, although electronic microscopy has been available for some time, achieving 1 nm resolution and even below, this technology cannot be applied to observe living cells and monitor their growth, because such cells will be killed. On the other hand, if optical lenses are used, the classical optical lower bound barrier of 200 nm does not allow us to apply our proposed mathematical formulation to accurately reveal the size and shapes of relatively smaller living cells. Fortunately, there was a recent break-through in optical microscopy of achieving super-resolution, by using fluorescent molecules, by three 2014 Nobel Laureates in Chemistry: Eric Betzig, Stephen Hell and William Moerner, that achieve optical super-resolution of 20 nm and even below (see \cite{wegel2016imaging}). The first generation of such super-resolution optical microscopy, including the STED (STimulated Emission Depletion) and SIM (Structural Illumination Microscopy) has been realized in the research laboratories, and already in production by such giant medical equipment companies as GE (see \cite{wegel2016imaging, GEpaidad}). By assuming that both $\x_{j,k} = \x_{j,k}(t)$ and $a_{j,k} = a_{j,k}(t)$ are functions of the time variable $t$, captured image sequence of cancer cells in discrete time facilitates monitoring the rate of cancer cell growth. Indeed, in the recent paper \cite{huang2016ultra} published last August, it was announced that 3-D optical microscopy at resolution of 10 nm - 20 nm wavelength can be realized at a fast speed, so that frame-by-frame 3-D videos of living cells could be constructed. Just imagine that if the growth of cancer cells can be identified at a early stage before they spread to other organs, then the recent research advancement in Oncology and Immune Engineering, in using the human immune system to kill cancer cells, could be applied to ``potential cancer patients” at a very early stage to save a lives!

\bhag{Final remarks}\label{remark}

The challenge in applying the mathematical tool developed in this paper is an effective and efficient computational scheme of the Hermite moments in \eref{nustardef} that are used as input to the point-mass isolation operator (PIO) introduced in \eref{hermitessodef}. In the companion paper \cite{chuigaussian}, a super-resolution approach based on Gaussian sums is introduced and developed to meet this challenge. The approach in \cite{chuigaussian} also provides a mathematical tool for separating real exponential sums, with applications to “isotope separation” in nuclear nhemistry, “echo cancellation” in signal processing, as well as other scientific and engineering problems.

%\bibliographystyle{abbrv}
%\bibliography{/Users/hrushikesh/Documents/hrushikesh/pctexfiles/hrushikesh}

\begin{thebibliography}{10}
\setlength{\itemsep}{0pt plus 0.3ex}
\bibitem{andrews_askey_roy}
G.~E. Andrews, R.~Askey, and R.~Roy.
\newblock {\em Special functions}, volume~71.
\newblock Cambridge university press, 1999.

\bibitem{askey1965mean}
R.~Askey and S.~Wainger.
\newblock Mean convergence of expansions in {L}aguerre and {H}ermite series.
\newblock {\em American Journal of Mathematics}, 87(3):695--708, 1965.

\bibitem{chuigaussian}
C.~K. Chui and H.~N. Mhaskar.
\newblock Multivariate super-resolution with application to decomposition of
  real exponential sums.
\newblock In preparation.

\bibitem{bspaper}
C.~K. Chui and H.~N. Mhaskar.
\newblock Signal decomposition and analysis via extraction of frequencies.
\newblock {\em Applied and Computational Harmonic Analysis}, 40(1):97--136,
  2016.

\bibitem{chung2015counting}
J.~Chung, X.~Ou, R.~P. Kulkarni, and C.~Yang.
\newblock Counting white blood cells from a blood smear using fourier
  ptychographic microscopy.
\newblock {\em PloS one}, 10(7):e0133489, 2015.

\bibitem{dziubanski1997triebel}
J.~Dziuba{\'n}ski.
\newblock {T}riebel-{L}izorkin spaces associated with {L}aguerre and {H}ermite
  expansions.
\newblock {\em Proceedings of the American Mathematical Society}, pages
  3547--3554, 1997.

\bibitem{epperson1997hermite}
J.~Epperson.
\newblock {H}ermite and {L}aguerre wave packet expansions.
\newblock {\em Studia Mathematica}, 126(3):199--217, 1997.

\bibitem{fang2017chip}
Y.~Fang, N.~Yu, R.~Wang, and D.~Su.
\newblock An on-chip instrument for white blood cells classification based on a
  lens-less shadow imaging technique.
\newblock {\em PloS one}, 12(3):e0174580, 2017.

\bibitem{frankbern}
F.~Filbir and H.~N. Mhaskar.
\newblock A quadrature formula for diffusion polynomials corresponding to a
  generalized heat kernel.
\newblock {\em Journal of Fourier Analysis and Applications}, 16(5):629--657,
  2010.

\bibitem{freud1972direct}
G.~Freud.
\newblock On direct and converse theorems in the theory of weighted polynomial
  approximation.
\newblock {\em Mathematische Zeitschrift}, 126(2):123--134, 1972.

\bibitem{huang2016ultra}
F.~Huang, G.~Sirinakis, E.~S. Allgeyer, L.~K. Schroeder, W.~C. Duim, E.~B.
  Kromann, T.~Phan, F.~E. Rivera-Molina, J.~R. Myers, I.~Irnov, et~al.
\newblock Ultra-high resolution 3d imaging of whole cells.
\newblock {\em Cell}, 166(4):1028--1040, 2016.

\bibitem{li2015red}
Q.~Li, M.~Zhou, H.~Liu, Y.~Wang, and F.~Guo.
\newblock Red blood cell count automation using microscopic hyperspectral
  imaging technology.
\newblock {\em Applied spectroscopy}, 69(12):1372--1380, 2015.

\bibitem{lou2016automatic}
J.~Lou, M.~Zhou, Q.~Li, C.~Yuan, and H.~Liu.
\newblock An automatic red blood cell counting method based on spectral images.
\newblock In {\em Image and Signal Processing, BioMedical Engineering and
  Informatics (CISP-BMEI), International Congress on}, pages 1391--1396. IEEE,
  2016.

\bibitem{tauberian}
H.~N. Mhaskar.
\newblock A unified framework for harmonic analysis of functions on directed
  graphs and changing data.
\newblock \textit{Applied and Computational Harmonic Analysis}, published
  online June 28, 2016.

\bibitem{mhasbk}
H.~N. Mhaskar.
\newblock {\em Introduction to the theory of weighted polynomial
  approximation}, volume~56.
\newblock World Scientific Singapore, 1996.

\bibitem{roy2014simple}
M.~Roy, G.~Jin, D.~Seo, M.-H. Nam, and S.~Seo.
\newblock A simple and low-cost device performing blood cell counting based on
  lens-free shadow imaging technique.
\newblock {\em Sensors and Actuators B: Chemical}, 201:321--328, 2014.

\bibitem{GEpaidad}
S.~Sanders.
\newblock Microscopy now: Getting the most from your imaging.
\newblock American Association for Advancement of Sciences, May 2015.
\newblock goo.gl/HkOtqI.

\bibitem{szego}
G.~Szeg\"o.
\newblock Orthogonal polynomials.
\newblock In {\em Colloquium publications/American mathematical society},
  volume~23. Providence, 1975.

\bibitem{wegel2016imaging}
E.~Wegel, A.~G{\"o}hler, B.~C. Lagerholm, A.~Wainman, S.~Uphoff, R.~Kaufmann,
  and I.~M. Dobbie.
\newblock Imaging cellular structures in super-resolution with sim, sted and
  localisation microscopy: A practical comparison.
\newblock {\em Scientific reports}, 6, 2016.

\end{thebibliography}

\end{document}